\documentclass[floatfix,showpacs,amsmath,twocolumn]{revtex4}
\usepackage{amssymb}
\def\be{\begin{eqnarray}}
\def\ee{\end{eqnarray}}
\def\bee{\begin{eqnarray*}}
\def\eee{\end{eqnarray*}}
\newtheorem{thm}{Theorem}
\newtheorem{cor}{Corollary}

\begin{document}

\title{Finite set of invariants to characterize local Clifford equivalence of stabilizer states}

\author{Maarten Van den Nest}
\email{maarten.vandennest@esat.kuleuven.ac.be} \author{Jeroen Dehaene}\author{Bart De Moor}
 \affiliation{Katholieke Universiteit Leuven, ESAT-SCD, Belgium.}
\date{\today}
\begin{abstract}
The classification of stabilizer states under local Clifford (LC) equivalence is of particular
importance in quantum error-correction and  measurement-based quantum computation. Two stabilizer
states are called LC equivalent if there exists a local Clifford operation which maps the first
state to the second. We present a finite set of invariants which completely characterizes the LC
equivalence class of any stabilizer state. Our invariants have simple descriptions within the
binary framework in which stabilizer states are usually described.
\end{abstract}

 \maketitle


Stabilizer states constitute a class of multipartite pure quantum states which is of considerable
interest in quantum information theory (QIT) and quantum computing (QC). A stabilizer state on $n$
qubits is defined as a simultaneous eigenvector of a maximal set of commuting observables in the
Pauli group, where the latter is the group generated by all $n$-fold tensor products of the Pauli
matrices and the identity. To name but two of their presently most prominent applications,
stabilizer states firstly appear as codewords of quantum stabilizer codes in the theory of quantum
error-correction \cite{Gott}, and, secondly, they are the resources for QC in a computation model
where only single-qubit measurements are performed (the one-way quantum computer \cite{1wayQC}). In
order to understand the role of stabilizer states in these and other QIT and QC tasks, the
mathematical and physical properties of these states have recently been studied in a number of
accounts \cite{entgraphstate, graphbriegel, localcliffgraph,invar_stab, alg_codes, val_bond}.
One central problem in this research is the classification of stabilizer states under local unitary
(LU) equivalence. Indeed, next to the natural relevance of this issue in the study of the
entanglement properties of stabilizer states, it is also of particular importance both in the
development of the one-way quantum computer and in the coding theoretic aspect of stabilizer
states. In this paper we consider a restricted version of LU equivalence of stabilizer states, in
that we consider only those local unitary operations which belong to the local Clifford group (LC).
The local Clifford group consists of all local unitary operators which map the Pauli group to
itself under conjugation. We will call two stabilizer states LC equivalent if there exists a local
Clifford operator which maps the first state to the second.

The restriction of considering LC equivalence rather than general LU equivalence has two main
motivations. Firstly, the problem of general LU equivalence is difficult and it is therefore
appropriate to first consider a more manageable subproblem, for which LC equivalence is a suitable
candidate - given the very explicit connections between stabilizers states, the Pauli group and the
(local) Clifford group. Secondly, this so-called 'subproblem' might in fact not be a subproblem at
all, as the natural question exists whether every two LU equivalent stabilizer states are
necessarily LC equivalent. Although this question is to date unanswered in general, our work
\cite{LU_LC}, following work performed in Ref. \cite{RainsMin2},  shows that this assertion is at
least true for some interesting subclasses of stabilizer states and a positive answer to this
question is widely believed.

In the following we study LC invariants of stabilizer states, i.e. functions in the entries of a
stabilizer state which take on equal values in LC equivalent states. We construct a finite
\emph{complete set} of invariants, which characterizes the LC equivalence class of any stabilizer
state. Such a study of LC invariants primarily serves to obtain a characterization of LC
equivalence classes of stabilizer states and to gain insight in the structure of these classes, and
to a lesser extent to obtain algorithms which can recognize LC equivalence of two given states.
Indeed, an efficient algorithm to recognize LC equivalence of stabilizer states has recently been
presented in our earlier work and there is little hope that equally efficient algorithms can be
constructed on the basis of invariants (see below). Therefore, we believe that the main merit of
the present result is, firstly, that it gives a \emph{finite} characterization of the LC
equivalence class of any stabilizer state and, secondly, that our invariants exhibit a very
transparent structure in terms of the stabilizer formalism.

Let us now start by recalling the basic notions concerning stabilizer states and the local Clifford
group. A stabilizer state $|\psi\rangle$ on $n$ qubits is the unique simultaneous eigenvector with
eigenvalue 1 of a set of $n$ commuting and independent observables in the Pauli group ${\cal G}_n$.
The latter consists of all operators of the form
$\alpha\sigma_{a_1}\otimes\dots\otimes\sigma_{a_n},$ where $\alpha\in\{ \pm1, \pm i\}$ is an
overall phase factor and $\sigma_{a_i}$ is either the $2\times 2$ identity matrix or one of the
Pauli matrices $\sigma_{x}$, $\sigma_y$, $\sigma_z$, for every $i=1, \dots, n$. The stabilizer
${\cal S}(\psi)$ of $|\psi\rangle$ is the set of all Pauli operators which have $|\psi\rangle$ as
an eigenvector with eigenvalue 1.
It is well known (see e.g. \cite{QCQI}) that the stabilizer formalism has an equivalent formulation
in terms of algebra over the field $\mathbb{F}_2=$ GF(2), where arithmetic is performed modulo 2.
The heart of this binary representation is an encoding of the Pauli matrices by pairs of bits,
writing $\sigma_0=\sigma_{00}$, $\sigma_x=\sigma_{01}$, $\sigma_z=\sigma_{10}$ and
$\sigma_y=\sigma_{11}$ and subsequently \be\sigma_{w_1w_1'}\otimes\dots\otimes\sigma_{w_n w_n'}=:
\sigma_{(w,w')}, \ee where $w:=(w_1, \dots, w_n),\ w':=(w_1', \dots, w_n')\in\mathbb{F}_2^{n}$ are
$n$-dimensional binary vectors. Note that the information about the $i$th qubit is distributed over
the $i$th components of the vectors $w$ and $w'$. Identifying every Pauli operator in an $n$-qubit
stabilizer ${\cal S}(\psi)$ with its $2n$-dimensional binary index vector, one can show that ${\cal
S}(\psi)$ corresponds to an $n$-dimensional self-dual linear subspace $S_{\psi}$ of
$\Bbb{F}_2^{2n}$; the self-duality of this subspace is with respect to a symplectic inner product
on $\Bbb{F}_2^{2n}$.

The Clifford group ${\cal C}_1$ on one qubit is the group of all $2\times 2$ unitary operators
which map $\sigma_u$ to $\alpha_u\sigma_{\pi(u)}$ under conjugation, where $u=x,y,z$, for some
$\alpha_u=\pm 1$ and  some permutation $\pi$ of $\{x,y,z\}$. The local Clifford group ${\cal
C}_n^l$ on $n$ qubits is the $n$-fold tensor product of ${\cal C}_1$ with itself. In the binary
stabilizer framework, local Clifford operations $U\in {\cal C}_n^l$ correspond to nonsingular
$2n\times 2n$ binary matrices $Q$ of the block form
\be Q = \left [ \begin{array}{cc} A&B\\
C&D \end{array}\right],\ee where the $n\times n$ blocks $A, B, C, D$ are diagonal
\cite{localcliffgraph}. We denote the diagonal entries of $A, B, C, D$ by
 $a_i$, $b_i$, $c_i$, $d_i$, respectively. The $n$  submatrices \be Q_i:=\left [
\begin{array}{cc} a_i & b_i
\\ c_i& d_i\end{array} \right ]\in GL(2,\Bbb{F}_2)\ee correspond to the tensor factors of $U$.
We denote the group of all such $Q$ by $C^l_n$. In the binary stabilizer framework, two $n$-qubit
stabilizer states $|\psi\rangle$, $|\psi'\rangle$ are LC equivalent if and only if there exists an
operator $Q\in C_n^l$ such that $QS_{\psi}=S_{\psi'}$, i.e., $Q$ maps the space $S_{\psi}$ to the
space $S_{\psi'}$.

Finally, let us recall the definition of \emph{support} in this context of stabilizers, which will
be of considerable importance in the following. The support supp$(v)$ of any vector $v\in
\Bbb{F}_2^{2n}$ is the set \be\label{supp} \mbox{supp}(v) &=& \{i\in \{1, \dots, n\}\ |\
(v_i,v_{n+i})\neq(0,0)\}\nonumber.\ee Note that supp$(v)$ contains exactly those $i\in\{1, \dots,
n\}$ such that the $i$th tensor factor of $\sigma_v$ differs from the identity.

We are now in a position to state the first result of this paper. In theorem 1 a finite set of
invariants which characterizes the LC equivalence class of any stabilizer state, is presented.
\begin{thm}
Let $|\psi\rangle$ be a stabilizer state on $n$ qubits. Let $r\in\Bbb{N}_0$ and consider subsets
$\omega^k$, $\omega^{kl}\subseteq\{1, \dots, n\}$ for every $ k,l\in\{1, \dots, r\}$ with $k<l$.
Denote $\Omega := (\omega^1, \omega^2, \dots, \omega^{12}, \omega^{13}, \dots)$ and let ${\cal
T}_{n,r}^{\Omega}(\psi)$ be the set consisting of all tuples $(v^1, \dots, v^r)\in
S_{\psi}\times\dots\times S_{\psi}$ satisfying \be\label{thm1} \mbox{ supp}(v^{k})= \omega^{k},\
\mbox{\ supp}(v^{k}+ v^{l})= \omega^{kl}.\ee Then (i) $|{\cal T}_{n,r}^{\Omega}(\psi)|$ is an LC
invariant and (ii) the LC equivalence class of $|\psi\rangle$ is completely determined by the
values of all invariants $|{\cal T}_{n,n}^{\Omega}(\psi)|$ (i.e. where $r=n$).
\end{thm}
{\it Proof: } Statement (i) is trivial. To prove (ii), let $|\psi\rangle$ and $|\psi'\rangle$ be
two stabilizer states on $n$ qubits such that $|{\cal T}_{n,n}^{\Omega}(\psi)| = |{\cal
T}_{n,n}^{\Omega}(\psi')|$ for every tuple $\Omega$ as defined above. We will show that this
implies that $|\psi\rangle$ and $|\psi'\rangle$ are LC equivalent. Fix a basis $\{v^1, \dots,
v^n\}$ of $S_{\psi}$ and define the tuple of sets $\Omega_*=(\omega_*^1,\dots, \omega_*^{12},
\dots)$  such that \be\label{thm1'} \omega_*^{k}=\mbox{ supp}(v^{k}) ,\ \omega_*^{kl}=\mbox{\
supp}(v^{k}+ v^{l}) ,\ee for every $k,l=1, \dots, n$ with $k<l$. It follows that $|{\cal
T}_{n,n}^{\Omega_*}(\psi)|$ is nonzero, and therefore $|{\cal T}_{n,n}^{\Omega_*}(\psi')|$ is
nonzero as well, since these numbers are equal by assumption. Therefore, there exist vectors $w^1,
\dots, w^n\in S_{\psi'}$ such that \be\label{thm1''} \mbox{ supp}(w^{k})= \omega_*^{k},\ \mbox{\
supp}(w^{k}+ w^{l})= \omega_*^{kl},\ee for every $k,l$.
Expressions (\ref{thm1'}) and ({\ref{thm1''}}) show that \be (v^{k}_{i}, v^{k}_{n+i}) = (0,0)
&\mbox{ iff }& (w^{k}_i,w^{k}_{n+i}) = (0,0)\nonumber\\ (v^{k}_i,v^{k}_{n+i}) =
(v^{l}_i,v^{l}_{n+i}) &\mbox{ iff }& (w^{k}_i,w^{k}_{n+i}) = (w^{l}_i,w^{l}_{n+i})\nonumber,\ee for
every $i,\ k,\ l=1, \dots, n$. This implies the existence of $n$
operators $Q_i\in GL(2, \mathbb{F}_2)$ such that \be Q_i \left[ \begin{array}{c} v^{k}_i\\
v^{k}_{n+i}\end{array}\right] = \left[ \begin{array}{c} w^{k}_i\\
w^{k}_{n+i}\end{array}\right]\ee for every $i,\ k =1, \dots, n$, and the operators $Q_i$
consequently constitute an operator $Q\in C_n^l$ such that $Qv^k = w^k$ for every $k=1, \dots, n$.
Therefore, the space $QS_{\psi}$ is spanned by the vectors $w^1, \dots, w^n$. Moreover, these
vectors are linearly independent, as they are the images of a linearly independent set under an
invertible linear transformation. Therefore, $\{w^1, \dots, w^n\}$ is a basis of $S_{\psi'}$ (since
the latter is an $n$-dimensional vector space) and we obtain $QS_{\psi} = S_{\psi'}$. This proves
the theorem. \hfill $\square$

Let us briefly discuss this result.  The invariants in theorem 1 are presented as cardinalities of
certain subsets of $S_{\psi}\times\dots\times S_{\psi}$, which are defined in terms of simple
constraints on the supports of their elements. Note that for $r=1$ these invariants simply count
the number of elements in the stabilizer with a prescribed support, i.e., for every
$\omega\subseteq\{1, \dots, n\}$ one has an invariant \be\label{equal}|\{ v\in S_{\psi}\ |\
\mbox{supp}(v)=\omega\}|. \ee These invariants are in fact 'local versions' of the so-called
\emph{weight distribution} of a stabilizer, a well known concept in quantum (and classical) coding
theory which is e.g. used to produce bounds on how good codes can be (see e.g. \cite{Shor_Lafl}).
Our invariants for $r\geq 2$ are in this respect generalizations of this well known notion.

While the structure of the invariants in theorem 1 is indeed very transparent, a direct calculation
of the numbers $|{\cal T}_{n,r}^{\Omega}(\psi)|$ is likely to be hard. It is in this context
interesting to consider the following variant of theorem 1, which presents an alternative complete
family of LC invariants:
\begin{thm}
Using the same notations as in theorem 1, let ${\cal V}_{n,r}^{\Omega}(\psi)$ be the vector space
consisting of all tuples $(v^1, \dots, v^r)\in S_{\psi}\times\dots\times S_{\psi}$ satisfying
\be\label{thm2} \mbox{ supp}(v^{k})\subseteq \omega^{k},\ \mbox{\ supp}(v^{k}+ v^{l})\subseteq
\omega^{kl}.\ee Then (i) the dimension of ${\cal V}_{n,r}^{\Omega}(\psi)$ is an LC invariant and
(ii) the LC equivalence class of $|\psi\rangle$ is completely determined by the values of all
invariants dim$({\cal V}_{n,n}^{\Omega}(\psi))$.
\end{thm}
{\it Proof: } Statement (i) is again trivial. We now prove (ii). Let $\omega\subseteq\{1, \dots,
n\}$ and define the functions $\delta_{\omega}, \epsilon_{\omega}: \Bbb{F}_2^{2n}\to \Bbb{C}$,
respectively,  by $\delta_{\omega}(w) = 1$ if supp$(w)=\omega$ and $\delta_{\omega}(w) = 0$
otherwise, and $\epsilon_{\omega}(w) = 1$ if supp$(w)\subseteq\omega$ and $\epsilon_{\omega}(w) =
0$ otherwise.
It is then straightforward to show the following relations: \be\label{relations1}
&&\epsilon_{\omega} = \sum_{\omega'\subseteq\omega} \delta_{\omega'}\nonumber\\
&&\delta_{\omega}=(-1)^{|\omega|}\sum_{\omega'\subseteq\omega}(-1)^{|\omega'|}\
\epsilon_{\omega'},\ee the first of which is trivial and the second of which can be verified using
straightforward combinatorics. Letting $\Omega$ be a tuple of sets
$\omega^k,\omega^{kl}\subseteq\{1, \dots, n\}$, with $k, l=1, \dots, r$ and $k<l$, it follows that
\be &&|{\cal T}_{n,r}^{\Omega}(\psi)| = \sum \left\{\prod_k \delta_{\omega^k}(v^k)\right\}
\left\{\prod_{k<l} \delta_{\omega^{kl}}(v^k+v^l)\right\}\nonumber\\
&&|{\cal V}_{n,r}^{\Omega}(\psi)| = \sum \left\{\prod_k \epsilon_{\omega^k}(v^k)\right\}
\left\{\prod_{k<l} \epsilon_{\omega^{kl}}(v^k+v^l)\right\}\nonumber,\ee where the sums run over all
$r$-tuples $(v^1, \dots, v^r)\in S_{\psi}^{\times r}$. Using (\ref{relations1}) one finds that
every invariant $|{\cal T}_{n,r}^{\Omega_0}(\cdot)|$ is a linear combination of the invariants
$\{|{\cal V}_{n,r}^{\Omega}(\cdot)|\}_{\Omega}$ with constant coefficients (and vice versa). The
proof then readily follows after applying theorem 1 and noting that the dimension of any vector
space ${\cal V}$ over $\mathbb{F}_2$ is given by dim$({\cal V})= \log_2 |{\cal V}|$. \hfill
$\square$

The main advantage of the invariants dim$({\cal V}_{n,r}^{\Omega}(\psi))$ over the $|{\cal
T}_{n,r}^{\Omega}(\psi)|$'s is that in the former case one is dealing with (dimensions of) vector
spaces, which are much more manageable from a computational point of view. Let us illustrate this
with a simple example: take $\omega\subseteq\{1, \dots, n\}$ as before and consider the invariant
\be\label{subset}|\{ v\in S_{\psi}\ |\ \mbox{supp}(v)\subseteq\omega\}|, \ee which is the
counterpart of (\ref{equal}) in theorem 2. This invariant can indeed be calculated efficiently as
follows: let $v^1, \dots, v^n$ be a basis of $S_{\psi}$ and let $S=[v^1| \dots| v^n]$ be the
$2n\times n$ binary matrix which has the vectors $v^j$ as its columns (the matrix S is generally
referred to as a \emph{generator matrix} of the stabilizer). Denoting by $S_i^T$ the $2\times n$
submatrix of $S$ which is obtained by assembling the $i$th and the $(n+i)$th row of $S$, for every
$i=1, \dots, n$, one readily verifies that the invariant (\ref{subset}) is equal to the corank
(i.e. the dimension of the kernel) over $\mathbb{F}_2$ of the matrix $S_{\omega}:=
(S^T_j)_{j\in\bar\omega}$, where $\bar\omega$ denotes the complement of $\omega$ in $\{1, \dots,
n\}$. Note that the calculation of the (co)rank of a matrix over any field indeed requires only a
polynomial number of operations in the dimensions of this matrix. This shows that the invariant
(\ref{subset}) can easily be calculated when a generator matrix of the stabilizer is known (note
that a stabilizer state in indeed usually presented in terms of a generator matrix). A
straightforward continuation of the above argument shows that, e.g. for $r=2$, an invariant
dim$({\cal V}_{n,2}^{\Omega}(\psi))$, where $\Omega=(\omega^1, \omega^2, \omega^{12})$ with
$\omega^1, \omega^2, \omega^{12}\subseteq\{1, \dots, n\}$, is equal to the corank
of the matrix \be \left[\begin{array}{cc} S_{\omega^1}& \cdot\\\cdot& S_{\omega^2}\\
S_{\omega^{12}} & S_{\omega^{12}}\end{array}\right],\ee and analogous results hold for arbitrary
degrees. This shows that the invariants dim$({\cal V}_{n,r}^{\Omega}(\psi))$ can indeed be
evaluated efficiently; moreover, the information needed to perform the calculation (i.e., the
matrices $S_{\omega}$) can easily be extracted from a generator matrix of the stabilizer.

There is in fact a close relationship between the LC invariants dim$({\cal
V}_{n,r}^{\Omega}(\psi))$ and the polynomial LU invariants  that we considered in
\cite{invar_stab}. In both cases, invariants correspond to dimensions of certain subspaces of
$S_{\psi}\times\dots\times S_{\psi}$ that are defined in terms of constraints on the supports of
their elements. In fact, if $r$ equals 1 or 2 then the LC invariants in theorems 1 and 2 are also
LU invariants; in other words, the invariants dim$({\cal V}_{n,r}^{\Omega}(\psi))$ and $|{\cal
T}_{n,r}^{\Omega}(\psi)|$ take on equal values on LU equivalent stabilizer states. This property is
well known (see e.g. \cite{RainsQWE}) for the invariants (\ref{equal}) and (\ref{subset}),
corresponding to $r=1$. In order to prove this assertion for $r=2$, let
$\Psi=|\psi\rangle\langle\psi|$ be (the projector associated with) a stabilizer state and take
$\Omega=(\omega^1, \omega^2, \omega^{12})$. Then one can readily verify that $|{\cal
V}_{n,2}^{\Omega}(\psi)|$ is, up to a multiplicative constant,  equal to \be \mbox{ Tr
}\left\{(\mbox{Tr}_{\bar\omega^1} \Psi)\ (\mbox{Tr}_{\bar\omega^2} \Psi)\
(\mbox{Tr}_{\bar\omega^{12}} \Psi)\right\},\ee which is manifestly invariant under the action of
local unitary operators (the operation Tr$_{\omega}$ denotes tracing out all qubits in the set
$\omega$). Moreover, this also shows that every $|{\cal T}_{n,2}^{\Omega}(\psi)|$ is an LU
invariant, as it can be written as a linear combination with constant coefficients of the $|{\cal
V}_{n,2}^{\Omega}(\psi)|$'s. Summarizing the current discussion, if $r$ equals 1 or 2 then the
invariants dim$({\cal V}_{n,r}^{\Omega}(\psi))$ and $|{\cal T}_{n,r}^{\Omega}(\psi)|$ take on equal
values on LU equivalent stabilizer states. It is clear that this result is very interesting in
relation with the problem whether LC equivalence and LU equivalence of stabilizer states are
identical notions.

Finally, we wish to note that the two main results in this paper are to be regarded as \emph{upper
bounds} on the maximal $r$ which needs to be considered in order to obtain a complete set of
invariants. Indeed, it is likely that only invariants corresponding to (substantially) lower $r$
need to be considered in order to recognize LC equivalence between two stabilizer states. One
indication for this conjecture is the proof of theorem 1, which shows that all the information of
the LC equivalence class of a given state $|\psi\rangle$ is in fact contained within a single
invariant $|{\cal T}_{n,n}^{\Omega_*}(\psi)|$. A second state $|\psi'\rangle$ is LC equivalent to
$|\psi\rangle$ if and only if \be |{\cal T}_{n,n}^{\Omega_*}(\psi)|=|{\cal
T}_{n,n}^{\Omega_*}(\psi')|, \ee (in fact, even $|{\cal T}_{n,n}^{\Omega_*}(\psi')|>0$ is
sufficient) and the information in all other invariants is in fact redundant. It is not unlikely
that there exist smaller complete lists of invariants (i.e. of smaller $r$) which exhibit less
redundancies. Secondly, our work shows examples of LC equivalence classes which are indeed
characterized by invariants of small $r$: e.g., any stabilizer state is LC equivalent to the GHZ
state on $n$ qubits if and only if the values of the invariants (\ref{subset}) coincide for both
states \cite{LU_LC}. We are therefore led to believe that the results in theorems 1 and 2 can in
principle be improved - if not for all stabilizer states then at least for some interesting
subclasses of states. It is in this context interesting to point out that it has been conjectured
and later disproved in a graph theoretical context that the list of LC invariants (\ref{subset})
are sufficient to characterize the LC equivalence class of all stabilizer states \cite{Bouchet,
entgraphstate}. Therefore, invariants where $r$ is at least 2 must be taken into account. It is to
date not clear whether also higher $r$ are to be considered.

Summarizing this paper, we have characterized LC equivalence classes of stabilizer states by means
of a finite set of invariants. Our invariants have simple descriptions within the binary stabilizer
framework, in that they are equal to dimensions of certain subspaces of $S_{\psi}\times\dots\times
S_{\psi}$. Furthermore, we discussed the link with existing families of LU invariants.

\begin{acknowledgments}

This research is supported by several funding agencies: Research Council KUL: GOA-Mefisto 666,
GOA-Ambiorics, several PhD/postdoc and fellow grants; Flemish Government: -   FWO: PhD/postdoc
grants, projects, G.0240.99 (multilinear algebra), G.0407.02 (support vector machines), G.0197.02
(power islands), G.0141.03 (Identification and cryptography), G.0491.03 (control for intensive care
glycemia), G.0120.03 (QIT), G.0452.04 (QC), G.0499.04 (robust SVM), research communities (ICCoS,
ANMMM, MLDM); -   AWI: Bil. Int. Collaboration Hungary/ Poland; -   IWT: PhD Grants, GBOU (McKnow)
Belgian Federal Government: Belgian Federal Science Policy Office: IUAP V-22 (Dynamical Systems and
Control: Computation, Identification and Modelling, 2002-2006), PODO-II (CP/01/40: TMS and
Sustainibility); EU: FP5-Quprodis;  ERNSI; Eureka 2063-IMPACT; Eureka 2419-FliTE; Contract
Research/agreements: ISMC/IPCOS, Data4s, TML, Elia, LMS, IPCOS, Mastercard; QUIPROCONE; QUPRODIS.
\end{acknowledgments}

\bibliographystyle{unsrt}
\bibliography{cliff_inv_graph_prl2}

\begin{thebibliography}{10}

\bibitem{Gott}
D.~Gottesman.
\newblock {\em Stabilizer codes and quantum error correction}.
\newblock PhD thesis, Caltech, 1997.
\newblock quant-ph/9705052.

\bibitem{1wayQC}
R.~Raussendorf, D.E. Browne, and H.J. Briegel.
\newblock Measurement-based quantum computation with cluster states.
\newblock {\em Phys. Rev. A}, 68:022312, 2003.
\newblock quant-ph/0301052.

\bibitem{entgraphstate}
M.~Hein, J.~Eisert, and H.J. Briegel.
\newblock Multi-party entanglement in graph states.
\newblock {\em Phys. Rev. A}, 69:062311, 2004.
\newblock quant-ph/0307130.

\bibitem{graphbriegel}
W.~D\"ur, H.~Aschauer, and H.J. Briegel.
\newblock Multiparticle entanglement purification for graph states.
\newblock {\em Phys. Rev. Lett.}, 91:107903, 2003.
\newblock quant-ph/0303087.

\bibitem{localcliffgraph}
M.~Van~den Nest, J.~Dehaene, and B.~De~moor.
\newblock Graphical description of the action of local clifford transformations
  on graph states.
\newblock {\em Phys. Rev. A}, 69:022316, 2004.
\newblock quant-ph/0308151.

\bibitem{invar_stab}
M.~Van~den Nest, J.~Dehaene, and B.~De~Moor.
\newblock Local invariants of stabilizer codes.
\newblock {\em Phys. Rev. A}, 70:032323, 2004.
\newblock quant-ph/0404106.

\bibitem{alg_codes}
M.~Van~den Nest, J.~Dehaene, and B.~De~Moor.
\newblock Efficient algorithm to recognize local clifford equivalence of graph
  states.
\newblock {\em Phys. Rev. A}, 70:034302, 2004.
\newblock quant-ph/0405023.

\bibitem{val_bond}
F.~Verstraete and J.I. Cirac.
\newblock Valence bond solids for quantum computation.
\newblock {\em Phys. Rev. A}, 70:060302, 2004.
\newblock quant-ph/0311130.

\bibitem{LU_LC}
M.~Van~den Nest, J.~Dehaene, and B.~De~moor.
\newblock Local unitary versus local clifford equivalence of stabilizer states.
\newblock {\em Phys. Rev. A}, 71:062323, 2005.
\newblock quant-ph/0411115.

\bibitem{RainsMin2}
E.~Rains.
\newblock Quantum codes of minimum distance two.
\newblock {\em IEEE Trans. Inform. Theory}, 45(1):266--271, 1999.
\newblock quant-ph/9704043.

\bibitem{QCQI}
I.~Chuang and M.~Nielsen.
\newblock {\em Quantum computation and quantum information}.
\newblock Cambridge University press, Cambridge, 2000.

\bibitem{Shor_Lafl}
P.~Shor and R.~Laflamme.
\newblock Quantum analog of the macwilliams identities for classical coding
  theory.
\newblock {\em Phys. Rev. Lett.}, 78(8):1600--1602, 1997.
\newblock quant-ph/9610040.

\bibitem{RainsQWE}
E.M. Rains.
\newblock Quantum weight enumerators.
\newblock {\em IEEE Trans. Inform. Theory}, 44(4):1388--1394, 1998.
\newblock quant-ph/9612015.

\bibitem{Bouchet}
A.~Bouchet.
\newblock Recognizing locally equivalent graphs.
\newblock {\em Discrete Math.}, 114(1-3):75--86, 1993.

\end{thebibliography}

\end{document}